\documentclass[a4paper,12pt]{article}
\usepackage{epsfig}
\usepackage[dvips,usenames]{color}
\usepackage{graphicx}

\newlength{\dinwidth}
\newlength{\dinmargin}
\newcommand{\spur}[1]{\not\! #1 \,}
\begin{document}
\title{ The rare $\bar{B}^0_{d} \to \phi\gamma$ decays in Standard Model and
as a probe of R-parity violation }

\bigskip

\author{Xinqiang Li,~ Gongru Lu,~ Rumin Wang, ~Y.D. Yang\footnote{ Corresponding
author. E-mail address: yangyd@henannu.edu.cn}
\\
{ \small \it Department of Physics, Henan Normal University,
Xinxiang, Henan 453002,  P.R. China}\\
}
\maketitle
\begin{picture}(0,0)
\put(305,290){\sf hep-ph/0305283}
\end{picture}

\bigskip\bigskip

\begin{abstract}
We present the first study of the rare annihilation decay
$\bar{B}^0_d \to \phi\gamma$ in the Standard Model. Using QCD
factorization formalism, we find ${\cal B}(\bar{B}^0_d
\to\phi\gamma )=3.6\times 10^{-12}$. The smallness of the decay
rate in the Standard Model make the decay a sensitive probe of new
physics contributions. As an example, we calculate the effects of
R-parity violating couplings. Within the available upper bounds
for $|\lambda^{''}_{i23}\lambda^{''}_{i12}|$ and
$|\lambda^{'}_{i32}\lambda^{'*}_{i12}|$, ${\cal B}( \bar{B}^0_d
\to\phi\gamma )$ could be enhanced to order of $10^{-9}\sim
10^{-8}$, which might be accessible at LHCB, B-TeV and the
planning super high luminosity B factories at KEK and SLAC.
\\
{\bf PACS Numbers 13.25.Hw,  12.60.Jv, 12.38.Bx, 12.15Mm}
\end{abstract}

\newpage
\section{Introduction}

\noindent It is known that flavor changing neutral currents(FCNC)
induced  rare B decays are very sensitive probes of new physics.
The GIM suppression of FCNC amplitude is absent in many new
physics scenarios beyond the Standard Model(SM), which could give
large enhancement of FCNC processes over the SM predictions. To
search such kind of signals is one of the most important goals of
B projects  BaBar, Belle, BTeV, and LHC-B. On the other hand, B
rare decays also serve as laboratory for hadronic dynamics. Due to
our poor knowledge for non-perturbative  QCD, predictions for many
interesting decays are always polluted by uncomfortable large
uncertainties, which have hindered us very much in  extracting
weak interaction information precisely from the available
measurements. The well known example is two-body charmless B
decays. It would be of great interesting to explore B rare decays
which are induced by FCNC currents as well as involve few hadronic
parameters.

To the end, we will study the pure penguin annihilation decay
$\bar{B}^0_{d} \to \phi\gamma$. Experimentally, $\bar{B}^0_d \to
\phi\gamma$  is very easy to be identified through the decay chain
$\bar{B}^0_d \to \phi\gamma\to( K^{-}K^{+})\gamma$, i.e, two
charge tracks and one energetic photon, and the detecting
efficiency should be high. However, if the decay rate is too rare,
it would be very difficult to pick out the signals of the decay
from its background, which comes from continuum ($e^+ e^- \to q
\bar q$ with q=u,d,s) events with high energy photons originating
from initial state radiation or $e^+ e^- \to (\pi^0 \eta)X$ with
$\pi^0 \eta\to\gamma\gamma$\cite{cleo00}. CLEO Collaboration had
performed a search for the decay, but found no evidence and put
the upper limit ${\cal B}(B^0_d \to \phi \gamma<0.33\times
10^{-5}$ at $90\%$ CL.  To our best knowledge, there is no
realistic theoretical study of  $\bar{B}^0_{d} \to \phi\gamma$ in
the literature. In this paper, we will investigate this decay. In
naive factorization approach, this decay involve the simple matrix
$\langle \phi|\bar{s}\gamma_{\mu}s|0\rangle$ and the same hadronic
matrix element $\langle
\gamma|\bar{q}\gamma_{\mu}(1-\gamma_{5})b|\bar{B_q}\rangle$ as the
radiative  leptonic decay which has been studied in
Ref.\cite{tmyan, sach, pirjol, kou} with different framework.
Beyond naive factorization, non-factorizable contribution should
be included. The QCD factorization framework\cite{BBNS}, which has
been developed recently  by Beneke $et~al.,$ is employed to
calculate ${\cal O}(\alpha_{s})$ nonfactorization contributions
arising  from exchanging hard gluon between the two color octet
currents $\bar{s}_{\alpha}\gamma_{\mu}(1\pm\gamma_{5})s_{\beta}$
and $\bar{d}_{\beta}\gamma^{\mu}(1-\gamma_{5})b_{\alpha}$. We find
${\cal B}(\bar{B}^0_{d}\to\phi\gamma)=3.6\times 10^{-12} $ in the
SM. The decay rate is too small to be observed at the running B
factories BaBar and BELLE. Any measurement of the decay  at BaBar
and BELLE would be the  evidence of activity of New Physics. As an
example, we treat it as a probe of R parity violating
couplings(RPV). Within the available lowest upper bound of RPV
couplings, it is found that the branching ratio of the decay could
be enhanced to $10^{-9}\sim 10^{-8}$, which might be measured at
LHC-B, BTeV and the planning super high luminosity B factories at
KEK and SLAC.

\section{ $\bar{B}^0_d \to\phi\gamma$ in the Standard Model}

The $\bar{B}^0_d \to\phi\gamma$ decay is illustrated schematically
in fig.1. which is dominated by the photon radiating from the
light quark in the B meson. Obviously, the amplitude of fig.1 is
suppressed by power of $\Lambda_{QCD}/m_b$ because $\phi$ meson
must be transversely polarized and B meson is heavy. The diagrams
with the photon radiating from the heavy b quark and the energetic
strange quarks of the $\phi$ meson are suppressed by power of
$\Lambda_{QCD}^2/m^2_b$, which will be neglected in this paper.
The situation is similar to that of the annihilation contributions
in $B\to K^* \gamma$ decays\cite{bosch}.

\begin{figure}[htbp]
\begin{center}
\scalebox{0.7}{\epsfig{file=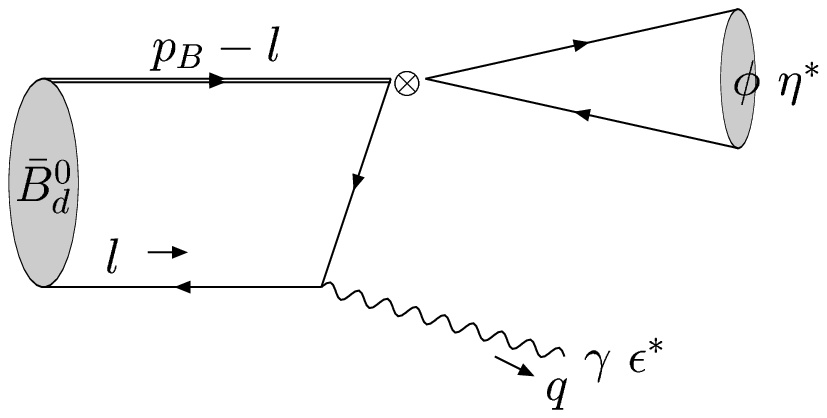}} \end{center}
\caption{\small The leading diagram  for $\bar{B}^0_d
\to\phi\gamma$. Radiation of the photon from the remaining three
quark lines are further suppressed by power of
$\Lambda_{QCD}/m_b$. The symbol  $\otimes$ denotes the insertion
of penguin operators $O_{3-10}$. }
\end{figure}

We begin our study with the effective  Hamiltonian in the SM
relevant to the decay \cite{buras}
\begin{equation}
\label{heff}
{\cal H}_{eff}
=-\frac{4G_{F}}{\sqrt{2}}
V_{tb} V_{td}^*
\left(\sum_{i=3}^{10}
C_{i}O_{i}
\right).
\end{equation}
For convenience, we list
 below the operators in  ${\cal H}_{eff}$ for $b \to d s \bar{s}$:
\begin{equation}\begin{array}{llllll}
O_3 & = & \bar d_\alpha\gamma^\mu L b_\alpha\cdot \bar
 s_\beta\gamma_\mu L s_\beta\ ,   &
O_4 & = & \bar d_\alpha\gamma^\mu L b_\beta\cdot \bar
s_\beta\gamma_\mu L s_\alpha\ , \\
O_5 & = & \bar d_\alpha\gamma^\mu L b_\alpha\cdot \bar
s_\beta\gamma_\mu R s_\beta\ ,   &
O_6 & = & \bar d_\alpha\gamma^\mu L b_\beta\cdot \bar
s_\beta\gamma_\mu R s_\alpha\ , \\
O_7 & = & \frac{3}{2}\bar d_\alpha\gamma^\mu L b_\alpha\cdot
e_{s}\bar s_\beta\gamma_\mu R s_\beta\ ,   &
O_8 & = & \frac{3}{2}\bar d_\alpha\gamma^\mu L b_\beta\cdot
e_{s}\bar s_\beta\gamma_\mu R s_\alpha\ , \\
O_9 & = & \frac{3}{2}\bar d_\alpha\gamma^\mu L b_\alpha\cdot
e_{s}\bar s_\beta\gamma_\mu L s_\beta\ ,   &
O_{10} & = & \frac{3}{2}\bar d_\alpha\gamma^\mu L b_\beta\cdot
e_{s}\bar s_\beta\gamma_\mu L s_\alpha.\\
\label{operators}
\end{array}
\end{equation}
Where $\alpha$ and $\beta$ are the $SU(3)$ color indices and
$L=(1 - \gamma_5)/2$, $R= (1 + \gamma_5)/2$.
The Wilson coefficients evaluated at $\mu=m_b$ scale are\cite{buras}
\begin{equation}
\begin{array}{llll}
    C_3=  0.014, &
    C_4= -0.035,&
    C_5=  0.009, &
    C_6= -0.041,\\
    C_7= -0.002/137,&
    C_8=  0.054/137,&
    C_9= -1.292/137,&
    C_{10}= 0.263/137,
\end{array}\label{ci}
\end{equation}
Using the effective Hamiltonian and naive factorization hypothesis, it
is easy to write down the amplitude for $\bar{B}^{0}_d \to\phi\gamma$
\begin{eqnarray}
A(\bar{B}^0_d \to\phi\gamma)&=&-\frac{G_{F}}{\sqrt{2}}V_{tb} V_{td}^*
\left[\left( C_3 +\frac{C_4}{N_c} +C_5 +\frac{C_6}{N_c} \right)
      -\frac{1}{2}\left( C_7 +\frac{C_8}{N_c}+ C_9 +\frac{C_{10}}{N_c} \right)
    \right]\nonumber \\
             &\times& \sqrt{4\pi\alpha_e}f_{\phi}m_{\phi} F_V
              \biggl\{ -\epsilon_{\mu\nu\rho\sigma}
              \eta^{*\mu}_{\perp}\epsilon^{*\nu}_{\perp}
             v^{\rho}q^{\sigma}
             +i\left[(\eta_{\perp}^{*}\cdot\epsilon^{*}_{\perp})
             ( v\cdot q)-(\eta^{*}_{\perp}\cdot q)
             (v\cdot\epsilon^{*}_{\perp})\right]\biggr\},
\end{eqnarray}
where $\eta^{*}_{\perp}$ and $\epsilon^{*}_{\perp}$ are transverse
polarization vectors of $\phi$ and photon respectively. The form
factor $F_V$ is defined by\cite{tmyan,sach,pirjol,kou}
\begin{equation}
\langle \gamma(\epsilon^{*},q){\mid} \bar{d}\gamma_{\mu}
(1-\gamma_{5})b {\mid}\bar{B}^0_d \rangle =
\sqrt{4\pi\alpha_e}\left[ -F_{V}
\epsilon_{\mu\nu\rho\sigma}\epsilon^{*\nu}v^{\rho}q^{\sigma}
+iF_{A}(\epsilon^{*}_{\mu} q{\cdot}v-q_{\mu}v{\cdot}\epsilon^{*})
\right].
\end{equation}
To leading power of ${\cal O}(1/m_{b})$, $F_V$ and $F_A$
read\cite{tmyan, sach}
\begin{equation}
F_{V}=F_{A}=\frac{Q_d f_{B}M_B}{ 2\sqrt{2}E_{\gamma} }\int dl_{+}\frac{\Phi_{B1}(l_{+})}{l_{+}}.
\end{equation}
The contributions of strong penguin operators arising from the
renormalization group evolution from the scale $\mu=M_W$ to
$\mu=m_b$ is very small due to the cancellations between them:
$C_{3}\simeq -C_{4}/3$ and $C_{5}\simeq -C_{6}/3$. Obviously the
amplitude is dominated by electro-weak penguin(EWP).

Now we can write down the helicity amplitude
\begin{eqnarray}
{\cal
M}_{+,+}&=&i\frac{G_F}{\sqrt{2}}V_{tb}V_{td}^{*}\sqrt{4\pi\alpha_e}F_{V}
f_{\phi}m_{\phi}M_{B}\left[ a_3 +a_5 -\frac{1}{2}(a_7 +a_9
)\right],\\ {\cal M}_{-,-}&=&0,
\end{eqnarray}
where $a_{i}=C_{i}+C_{i+1}/N_{C}$. It is interesting to note  that
the $\phi$ meson and the photon in the decay are right-handed
polarized in the SM. It is also easy to realize that the decay is
very rare because of helicity suppression as well as small
$V_{td}$, $f_{B}$ and $f_{\phi}$. Using
$f_{\phi}=254MeV$\cite{ball}, $f_{B}=180MeV$, $|V_{td}|=0.008$,
$N_C =3$ and $\lambda_{B}^{-1}=\int
dl_{+}\Phi_{B1}(l_{+})/l_{+}=(0.35GeV)^{-1}$\cite{sach,BBNS}, we
get
\begin{equation}
{\cal B}(\bar{B}^0_d \to\phi\gamma )=3.5\times 10^{-13}.
\end{equation}

In the above calculations, non-factorizable contributions are
neglected. However, the leading non-factorizable diagrams in Fig.2
should be taken into account. For this purpose, the QCD
factorization framework\cite{BBNS} invented recently by Beneke,
Buchalla, Neubert and Sachrajda is very suitable. The framework
incorporate many important theoretical aspects of QCD like color
transparency, heavy quark limit and hard-scattering which allows
us to calculate nonfactorizable contributions systematically.

To calculate the nonfactorizable diagrams as depicted by Fig.2, we
take photon and $\phi$ meson flying along $n_{-}=(1,0,0,-1)$ and
$n_{+}=(1,0,0,1)$ directions respectively. We need the
two-particle light-cone projector for B meson and $\phi$ meson
\begin{eqnarray}
{\cal M}^B_{\alpha\beta}&=& \frac{i}{4 N_c}f_{B}M_{B}
\biggl\{(1+\spur{v})\gamma_5
\left[\Phi_{B1}(l_{+})+\spur{n_{-}}\Phi_{B2}(l_{+}) \right]
\biggr\}_{\alpha\beta}, \\
 {\cal M}^{\phi}_{\perp\delta\gamma}&=&-\frac{f^{\perp}_{\phi}m_{\phi}}{4N_c}
\biggl\{\spur{\epsilon}^*_{\perp}g^{(v)}_{\perp}(u)+i\epsilon_{\mu\nu\rho\sigma}
\epsilon^{*\nu}_{\perp}n^{\rho}_{+}n^{\sigma}_{-}
\gamma^{\mu}\gamma_{5}\frac{g^{(a)\prime}_{\perp}(u)}{8}\biggr\}_{\delta\gamma},
\end{eqnarray}
which encode the relevant non-perturbative bound state dynamics of
the initial B meson and the final $\phi$ meson. $\Phi_{B1}(l_{+})$
and $\Phi_{B2}(l_{+})$ are the leading twist light-cone
distribution function of B meson\cite{grozin}.
$g^{(v)}_{\perp}(u)$ and $g^{(a)}_{\perp}(u)$ are twist-3
distribution amplitude of $\phi$ meson\cite{ball},
$g^{(a)\prime}_{\perp}(u)=dg^{(a)}_{\perp}(u)/du$. The detail
discussions on these projectors could be found in
Refs.\cite{ball,grozin,benekef}.

\begin{figure}[htbp]
\begin{center}\scalebox{0.7}{\epsfig{file=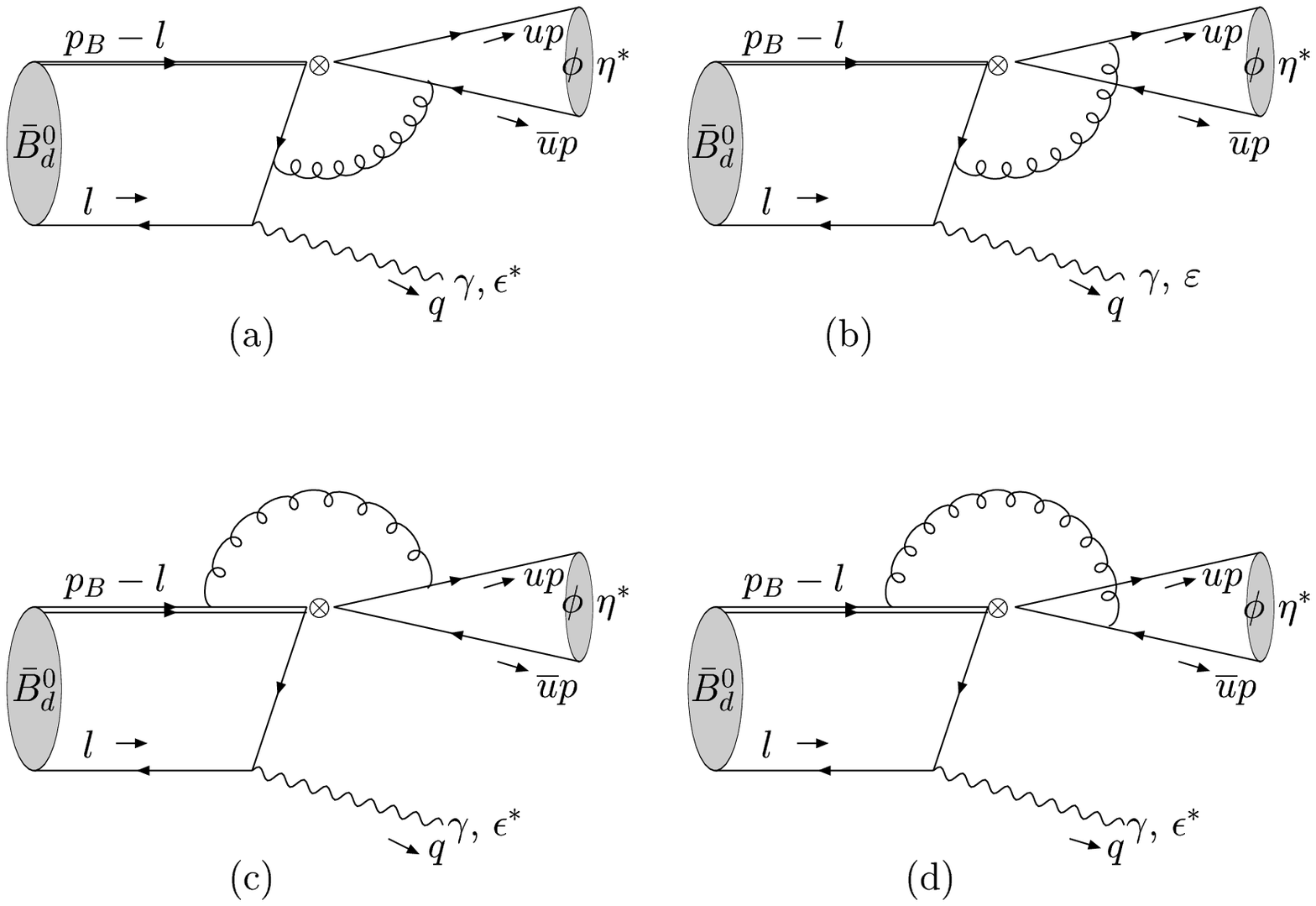}}
\end{center}
 \caption{\small Nonfactorizable diagrams for  $\bar{B}^0_d
\to\phi\gamma$. Other diagrams with the photon radiating from the
heavy b quark and  the energetic quark lines are suppressed.}
\end{figure}
The amplitude for $\bar{B}^0_d \to \phi\gamma$ can be presented by
the schematic formula
\begin{eqnarray}
\langle \phi\gamma|{\cal H}_{eff}|B\rangle&=&
-\frac{G_{F}}{\sqrt{2}}V_{tb} V_{td}^*
\left[\left( a^{\prime}_3  +a^{\prime}_5  \right)
      -\frac{1}{2}\left( a^{\prime}_7 + a^{\prime}_9  \right)
    \right]\nonumber \\
&\times& \sqrt{4\pi\alpha_e}f_{\phi}m_{\phi} F_V \biggl\{
-\epsilon_{\mu\nu\rho\sigma}
\eta^{*\mu}_{\perp}\epsilon^{*\nu}_{\perp}
 v^{\rho}q^{\sigma}
             +i\left[( \eta_{\perp}^{*}\cdot\epsilon^{*}_{\perp} v{\cdot}q )
             -(\eta^{*}_{\perp}\cdot q)
             (v\cdot\epsilon^{*}_{\perp})\right]
              \biggr\}.
\end{eqnarray}
The ${\cal O}(\alpha_s )$ corrections are summarized in
$a_{i}^{\prime}$'s which are calculated to be
\begin{eqnarray}
a^{\prime}_3&=&a_3
+\frac{\alpha_s}{4\pi}\frac{C_F}{N_C}\frac{f_{\phi}^{\perp}}{f_{\phi}}C_{4}F_{1},\\
a^{\prime}_5&=&a_5
+\frac{\alpha_s}{4\pi}\frac{C_F}{N_C}\frac{f_{\phi}^{\perp}}{f_{\phi}}C_{6}F_{2},\\
a^{\prime}_7&=&a_7
+\frac{\alpha_s}{4\pi}\frac{C_F}{N_C}\frac{f_{\phi}^{\perp}}{f_{\phi}}C_{8}F_{2},\\
a^{\prime}_9&=&a_9
+\frac{\alpha_s}{4\pi}\frac{C_F}{N_C}\frac{f_{\phi}^{\perp}}{f_{\phi}}C_{10}F_{1},
\end{eqnarray}
where $F_{1,2}$ arise  from one gluon exchange between the two
currents of color-octet penguin operator ${\cal O}_{4,6,8,10}$ as
shown by fig.2.
\begin{eqnarray}
F_{1}&=&\int^1_0 du\left(\frac{g_{\perp}^{(a)\prime}(u)}{4}
-g^{(v)}_{\perp}(u) \right) \left[ -14-3i\pi-12\ln\frac{\mu}{m_b}
\right. \nonumber\\ &&\left. +\left( 5+\frac{u}{1-u}\ln u \right)
-\frac{\pi^2}{3}+2Li_{2}(\frac{u-1}{u})\right],\\ F_{2}&=&\int^1_0
du\left( g^{(v)}_{\perp}(u)+\frac{g_{\perp}^{(a)\prime}(u)}{4}
                   \right)
\left[ -14-3i\pi-12\ln\frac{\mu}{m_b} \right. \nonumber\\ &&\left.
+\left( 5+\frac{u}{1-u}\ln u \right)
-\frac{\pi^2}{3}+2Li_{2}(\frac{u-1}{u})\right].
\end{eqnarray}
 In the calculation, the $\overline{MS}$ renormalization scheme is used.
 We have neglected the small effect of box diagrams and the
 diagrams with photon  radiating from energetic strange quarks,
which are further suppressed by $\Lambda_{QCD}/M_{B}$. We also
have neglect $l^2_{+}$ terms entered in the loop calculation which
are the higher twist effect, in this way, the integral involved
$\Phi_{B2}(l_{+})$ absents and the remain integrals are related to
the form factor $F_V$.

Including ${\cal O}(\alpha_{s})$ contributions, the helicity
amplitude is
\begin{eqnarray}
{\cal
M}_{++}&=&i\frac{G_F}{\sqrt{2}}V_{tb}V_{td}^{*}\sqrt{4\pi\alpha_e}F_{V}
f_{\phi}m_{\phi}M_{B}\left[ a^{\prime}_3 +a^{\prime}_5
-\frac{1}{2}(a^{\prime}_7 +a^{\prime}_9 )\right],\nonumber\\ {\cal
M}_{--}&=&0. \label{am1}
\end{eqnarray}
To give numerical result, we take $\mu=m_b$ and
$f_{\phi}^{\perp}=215MeV$\cite{ball}. The branching ratio is
estimated to be
\begin{equation}
{\cal B}(\bar{B}^0_d \to\phi\gamma )=3.6\times 10^{-12}.
\end{equation}
It indicates that the decay is too rare to be measured at the
running B factories BaBar and BELLE. However, the decay may be
accessible at LHCB and the planning super high luminosity B
factories at KEK and SLAC. Furthermore,  it could be enhanced by
New Physics,  and the enhancement might be  large enough for being
measured at these facilities.

\section{$\bar{B}^0_d \to\phi\gamma$  as a probe of R parity violation}

In the minimal supersymmetric standard model( MSSM ) \cite{MSSM} a
discrete symmetry called R-parity is invoked to forbid gauge
invariant lepton and baryon number violating operators. The
R-parity of a particle is given by\cite{farrar} $R_p
=(-1)^{L+2S+3B}$, where L and B are lepton and baryon numbers, and
S is the spin.  However there is no deep theoretical motivation
for imposing R-parity  and it is interesting to explore  the
phenomenology of R-parity violation\cite{Rgroup}.

We start our exploration from the R-parity violating
superpotential
\begin{equation}
W_{\spur{R}}=\frac{1}{2}\lambda LLE^{c}+\lambda'
LQD^{c}+\frac{1}{2}\lambda'' U^{c} D^{c} D^{c}.
\end{equation}
We are interested in the $\lambda'$ and $\lambda''$  terms since
they are relevant to the process $b\to ds\bar{s}$. Writing down
all indices explicitly, it reads
\begin{equation}
W_{\spur{R}}=\varepsilon^{ab}\delta^{\alpha\beta}\lambda^{'}_{ijk}L_{ia}
Q_{jb\alpha}D^{c}_{k\beta}
+\frac{1}{2}\varepsilon^{\alpha\beta\gamma} \left[
\lambda''_{i[jk]} U^{c}_{i\alpha} D^{c}_{j\beta} D^{c}_{k\gamma}
\right],
\end{equation}
where $U^{c}_{i\alpha}$, $D^{c}_{j\beta}$ and  $D^{c}_{k\gamma}$ are the
superfields of the right handed quarks and/or squarks respectively, the superscript
c denotes charge conjugation. i, j and k are generation indices and $\alpha$,
$\beta$ and $\gamma$ are $SU(3)$ color triplet indices. It follows from the
antisymmetry of $\varepsilon^{\alpha\beta\gamma}$ that $\lambda''_{i[jk]}$
is antisymmetric in the last two indices.

In terms of four components Dirac spinor, from $W_{\spur{R}}$ we can read
\begin{equation}
{\cal L}_{eff}^{\spur{R}}=\frac{1}{2} \varepsilon^{\alpha\beta\gamma} \left[
\lambda''_{ijk} \tilde{u}_{Ri\alpha}
\left(\bar{d^c}_{j\beta} R d_{k\gamma}-\{j\leftrightarrow k\}\right)
\right]
 +\lambda'_{ijk} \tilde{\nu}_{Li}\bar{d}_{k} L d_{j}
+h.c.
\end{equation}
 From ${\cal L}_{eff}$, we get the effective Hamiltonian for $b\to ds\bar{s}$
\begin{eqnarray}
{\cal H}_{\spur{R}}=\frac{1}{m^2_{\tilde{u}_i} }
 \varepsilon^{\alpha\beta\gamma} \varepsilon^{\alpha\beta'\gamma'}
\left[ \lambda''_{i23} \lambda''^{*}_{i12}
       ( \bar{s}^{c}_{\beta} R b_{\gamma} )
       ( \bar{s}_{\gamma'} L d^c_{\beta'} )
\right.
&+&
 \lambda''_{i23} \lambda''^{*}_{i21}
       ( \bar{s}^{c}_{\beta} R b_{\gamma} )
       ( \bar{d}_{\gamma'} L s^c_{\beta'} )
\nonumber \\
+ \lambda''_{i32} \lambda''^{*}_{i12}
       ( \bar{b}^{c}_{\beta} R s_{\gamma} )
       ( \bar{s}_{\gamma'} L d^c_{\beta'} )
&+& \left. \lambda''_{i32} \lambda''^{*}_{i21}
       ( \bar{b}^{c}_{\beta} R s_{\gamma} )
       ( \bar{d}_{\gamma'} L s^c_{\beta'} )
\right ] \nonumber\\
+\frac{1}{m^2_{\tilde{\nu}_i} }
\left[ \lambda'_{i31}\lambda'^{*}_{i22}(\bar{s}_{\alpha} R s_{\alpha})
(\bar{d}_{\beta} L b_{\beta})
\right.
&+& \lambda'^{*}_{i13}\lambda'_{i22}(\bar{s}_{\alpha} L s_{\alpha})
(\bar{d}_{\beta} R b_{\beta})
 \nonumber \\
 +\lambda'_{i32}\lambda'^{*}_{i12}(\bar{d}_{\alpha} R s_{\alpha})
(\bar{s}_{\beta} L b_{\beta})
    &+&
\left. \lambda'^{*}_{i23}\lambda'_{i21}(\bar{d}_{\alpha} L s_{\alpha})
 (\bar{s}_{\beta} R b_{\beta})
\right].
\end{eqnarray}
Contracting $\varepsilon^{\alpha\beta\gamma}
\varepsilon^{\alpha\beta'\gamma'}$ and performing Fierz
transformations, we get
\begin{eqnarray}
{\cal H}_{\spur{R}}(b\to ds\bar{s}) &=&
 -\frac{2}{ m^2_{\tilde{u}_i} }
\lambda''_{i23} \lambda''^{*}_{i12}
 \left[
       ( \bar{s}_{\beta}\gamma_{\mu} R s_{\beta} )
       ( \bar{d}_{\gamma}\gamma^{\mu} R b_{\gamma} )
   - ( \bar{s}_{\beta}\gamma_{\mu} R s_{\gamma} )
       ( \bar{d}_{\gamma}\gamma^{\mu} R b_{\beta} )
\right] \nonumber\\
&&-\frac{1}{2m^2_{\tilde{\nu}_i}}
 \left[
 \lambda'_{i31}\lambda'^{*}_{i22}(\bar{s}_{\alpha} \gamma_{\mu} L b_{\beta})
(\bar{d}_{\beta} \gamma^{\mu} R s_{\alpha})
+\lambda'^{*}_{i13}\lambda'_{i22}(\bar{s}_{\alpha} \gamma_{\mu} R b_{\beta})
(\bar{d}_{\beta} \gamma^{\mu} L s_{\alpha})
   \right. \nonumber\\
&& \left.
 +\lambda'_{i32}\lambda'^{*}_{i12}(\bar{d}_{\alpha} \gamma_{\mu} L b_{\beta})
(\bar{s}_{\beta} \gamma^{\mu} R s_{\alpha})
+\lambda'^{*}_{i23}\lambda'_{i21}(\bar{d}_{\alpha} \gamma_{\mu} R b_{\beta})
(\bar{s}_{\beta} \gamma^{\mu} L s_{\alpha})\right],
\end{eqnarray}
 where we have used  the relation
\begin{equation}
 \left( \bar{q}^{c}\gamma_{\mu} L q'^{c} \right)
= -\left( \bar{q'}\gamma_{\mu} R q \right).
 \end{equation}
The effective Hamiltonian is near to the idea form. Using
renormalization group to run it from sfermion mass scale
$m_{\tilde{f}_i}$ (100GeV assumed ) down to $m_b$ scale, it reads
\begin{eqnarray}
{\cal H}_{\spur{R}}(b\to ds\bar{s}) &=&
 -\frac{2}{m^2_{\tilde{u}_i} } \eta^{-4/\beta_0 }
\lambda''_{i23} \lambda''^{*}_{i12}
 \left[
       ( \bar{s}_{\beta}\gamma_{\mu} R s_{\beta} )
       ( \bar{d}_{\gamma}\gamma^{\mu} R b_{\gamma} )
   - ( \bar{s}_{\beta}\gamma_{\mu} R s_{\gamma} )
    ( \bar{d}_{\gamma}\gamma^{\mu} R b_{\beta} )
\right] \nonumber\\
&&-\frac{1}{2m^2_{\tilde{\nu}_i}}\eta^{-8/\beta_0 }
 \left[
 \lambda'_{i31}\lambda'^{*}_{i22}(\bar{s}_{\alpha} \gamma_{\mu} L b_{\beta})
(\bar{d}_{\beta} \gamma^{\mu} R s_{\alpha})
+\lambda'^{*}_{i13}\lambda'_{i22}(\bar{s}_{\alpha} \gamma_{\mu} R b_{\beta})
(\bar{d}_{\beta} \gamma^{\mu} L s_{\alpha})
   \right. \nonumber\\
&& \left.
 +\lambda'_{i32}\lambda'^{*}_{i12}(\bar{d}_{\alpha} \gamma_{\mu} L b_{\beta})
(\bar{s}_{\beta} \gamma^{\mu} R s_{\alpha})
+\lambda'^{*}_{i23}\lambda'_{i21}(\bar{d}_{\alpha} \gamma_{\mu} R b_{\beta})
(\bar{s}_{\beta} \gamma^{\mu} L s_{\alpha})\right],
\end{eqnarray}
with $\eta=\frac{\alpha_{s}(m_{\tilde{f}_i})}{\alpha_{s}(m_b )}$ and
$\beta_0 =11-\frac{2}{3}n_f $.

Now we are ready to write down R-parity violation contributions in
$\bar{B}^0_d  \to \phi\gamma$ decays
\begin{eqnarray}
{\cal M}^{\spur{R}}(\bar{B}^0_d   \to \phi\gamma)&=& \sqrt{4 \pi
\alpha_e } F_{V} f_{\phi}m_{\phi} \left[
i(\eta^{*}_{\perp}\cdot\epsilon^{*}_{\perp})(v\cdot q
)+\epsilon_{\mu\nu\alpha\beta}\eta_{\perp}^{*\mu}
\epsilon_{\perp}^{*\nu}v^{\alpha}q^{\beta} \right] \nonumber \\
&&\times \biggl\{
 \frac{1}{2 m^2_{\tilde{u}_i}} \lambda''_{i23}
\lambda''^{*}_{i12}\eta^{-4/\beta_0 } \left(
1-\frac{1}{N_{C}}-\frac{\alpha_s }{4\pi} \frac{C_F}{N_c} F_{1}
\right) \biggr. \nonumber \\ && \biggl.
 +\frac{1}{8 m^2_{\tilde{\nu}_i}} \lambda'_{i21}
\lambda'^{*}_{i23}\eta^{-8/\beta_0 } \left(
\frac{1}{N_{C}}+\frac{\alpha_s }{4\pi} \frac{C_F}{N_c} F_{2}
\right) \biggr\} \nonumber \\
 &-&\sqrt{4 \pi \alpha_e } F_{V}
f_{\phi}m_{\phi} \left[
i(\eta^{*}_{\perp}\cdot\epsilon^{*}_{\perp})(v\cdot q)
-\epsilon_{\mu\nu\alpha\beta}\eta_{\perp}^{*\mu}
\epsilon_{\perp}^{*\nu}v^{\alpha}q^{\beta} \right] \nonumber \\
&&\times \frac{1}{8 m^2_{\tilde{\nu}_i}} \lambda'_{i32}
\lambda'^{*}_{i12}\eta^{-8/\beta_0 } \left(
\frac{1}{N_{C}}+\frac{\alpha_s }{4\pi} \frac{C_F}{N_c} F_{1}
\right).
\end{eqnarray}
We note that only $u$ channel  squarks and sneutrino mediated
terms contribute because of $\langle
\gamma|\bar{d}(1\pm\gamma_{5})b|\bar{B}^0_d \rangle=0$.
Immediately, we derive
\begin{eqnarray}
{\cal M}^{\spur{R}}_{++}&=&iM_{B}\sqrt{4 \pi \alpha_e } F_{V}
f_{\phi}m_{\phi} \frac{1}{8 m^2_{\tilde{\nu}_i}} \lambda'_{i32}
\lambda'^{*}_{i12}\eta^{-8/\beta_0 } \left(
\frac{1}{N_{C}}+\frac{\alpha_s }{4\pi} \frac{C_F}{N_c} F_{1}
\right),
\\
 {\cal M}^{\spur{R}}_{--}&=&-iM_{B}\sqrt{4 \pi \alpha_e } F_{V}
f_{\phi}m_{\phi} \biggl\{ \frac{1}{2 m^2_{\tilde{u}_i}}
\lambda''_{i23} \lambda''^{*}_{i12}\eta^{-4/\beta_0 } \left(
1-\frac{1}{N_{C}}-\frac{\alpha_s }{4\pi} \frac{C_F}{N_c} F_{1}
\right) \biggr. \nonumber \\ && \biggl.
 +\frac{1}{8 m^2_{\tilde{\nu}_i}}
\lambda'_{i21} \lambda'^{*}_{i23}\eta^{-8/\beta_0 } \left(
\frac{1}{N_{C}}+\frac{\alpha_s }{4\pi} \frac{C_F}{N_c} F_{2}
\right) \biggr\}.
\end{eqnarray}

\begin{figure}[htbp]
\begin{center}
\scalebox{0.7}{\epsfig{file=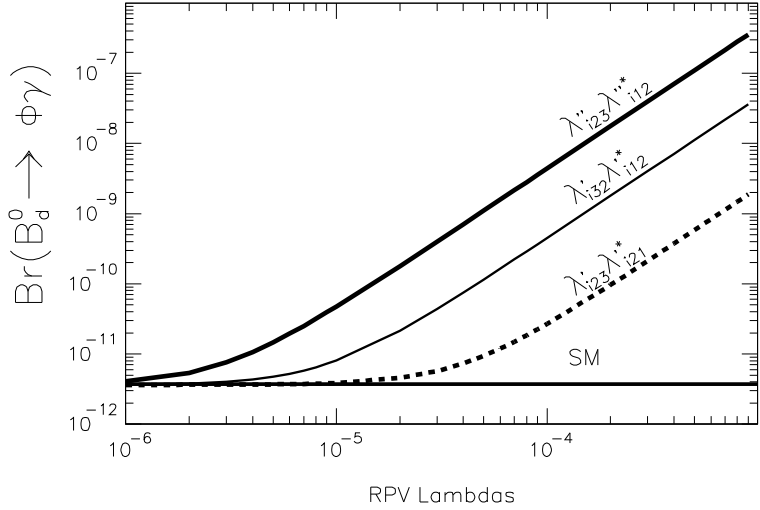}}
\end{center}
 \caption{\small The branching ratio of $\bar{B}^0_d \to\phi\gamma$ as a function of
 R-parity violation couplings $|\lambda''_{i23} \lambda''^{*}_{i12}| $(the thick solid
 curve), $|\lambda'_{i32} \lambda'^{*}_{i12}| $ (the thin solid curve) and
  $|\lambda'_{i21} \lambda''^{*}_{i23}| $(the dash curve) respectively.
  The  horizontal line is  the SM prediction.}
\end{figure}

To present numerical results, we assume that only one sfermion
contribute one time and universal mass 100GeV for sfermions
$\tilde{u}_{Ri}$. Our results for R parity violation contributions
are summarized in fig.3. We note that the decay is very sensitive
to R parity violating couplings. Within the available lowest upper
bounds for $|\lambda''_{i23} \lambda''^*_{i21}|$ \cite{yangeilam,
xghe, dreiner},
\begin{equation}
| \lambda^{''}_{i23}\lambda^{''*}_{i21}|<6\times10^{-5}
\left(\frac{m_{\tilde{u}_{Ri}}}{100}\right)^2 , ~~ |
\lambda^{'}_{i32}\lambda^{'*}_{i12}|<4\times10^{-4}
\left(\frac{m_{\tilde{\nu}_{Li}}}{100}\right)^2 ,
\end{equation}
${\cal B}(\bar{B}^0_{d} \to \phi\gamma)$ could be enhanced to
$10^{-9}\sim 10^{-8}$ which may  be inaccessible at BELLE and
BaBar. However, it is large enough for  LHCb, BTeV and the
planning super high luminosity B factories at KEK and SLAC.

\section{Conclusion}
We have studied pure penguin radiative annihilation process
$\bar{B}^{0}_{d}\to\phi\gamma$ by using QCD factorization for the
hadronic dynamics. We find that non-factorizable contributions are
larger than factorizable contributions in
$\bar{B}^{0}_d\to\phi\gamma$ decays. We estimate ${\cal
B}(\bar{B}^0_{d} \to \phi\gamma )=3.6\times10^{-12}$ in the SM.
The smallness of these decays in the SM makes it a sensitive probe
of flavor physics beyond the SM. To explore the potential, we have
estimated the contribution of RPV couplings to this decay. We have
found that the decay $\bar{B}^{0}_{d}\to\phi\gamma$ is very
sensitive to the $\lambda'' U^{c} D^{c} D^{c}$ and  $\lambda' LQD$
terms in the R-parity violating superpotential. Within the upper
limits for $\lambda''_{i23} \lambda''^{*}_{i21}$ and
$\lambda'_{i32} \lambda'^{*}_{i12}$ , it is found that ${\cal
B}(\bar{B}^0_{d} \to \phi\gamma)$ could be enhanced to order of
$10^{-9}\sim 10^{-8}$. We also note that the $\phi$ meson and the
photon are right-handed polarized in the SM, but they can be
left-handed polarized in RPV supersymmetry.  We find that
$\lambda'_{i32} \lambda'^{*}_{i12}$ RPV couplings contribute to
the right-handed polarized magnitude while $\lambda''_{i23}
\lambda''^{*}_{i12}$ and $\lambda'_{i21} \lambda'^{*}_{i23}$ RPV
couplings responsible to the left-handed polarized magnitude.
Theoretically, measuring the polarization of  photon (or the
$\phi$ meson ) can be used to probe the relative strength of RPV
couplings. However, the interesting phenomena is faded by the
small branching ratio of the exotic decay. In literature, there
are interesting proposals for probing new physics by measuring the
photon polarization in radiative decays $B\to K^*_{i}
\gamma$\cite{gronau}. It always needs large amount of experimental
data. For the more exotic decay $\bar{B}^0_d \to \phi\gamma$, this
phenomena will be remained academic unless the continuum
background could be suppressed efficiently in data analyzing at
the future B facilities. The interesting decay may be too rare to
be accessible at Babar and Belle. However, the decay rate might be
studied by LHCB at CERN, BTeV at Fermilab and the planning super
high luminosity B factories at KEK and SLAC to probe new physics.

\section*{Acknowledgments}
One of the authors(Y.D) would greatly appreciate Anirban Kundu for
helpful discussions on RPV contributions. We thank D.E. Jaffe and
H.B. Li for pointing the Ref.\cite{cleo00} to us and useful
comments on the continuum background.  Y.D is supported by the
Henan Provincial Science Foundation for Prominent Young Scientists
under the contract 0312001700. This work is supported in part by
National Science Foundation of China under the contracts 19805015
and 1001750.


\end{document}